**Fundamental nature of the self-field critical current in superconductors***

Evgeny F. Talantsev[1] and Jeffery L. Tallon[2]

[1]M.N. Miheev Institute of Metal Physics, Ural Branch, Russian Academy of Sciences, 18, S. Kovalevskoy St., Ekaterinburg, 620108, Russia

[2]Robinson Research Institute, Victoria University of Wellington, 69 Gracefield Road, Lower Hutt, 5040, New Zealand

**The ability to conduct electric current without dissipating energy is a property of superconductors that is used in magnetic systems utilized in healthcare, natural sciences, and ongoing global projects in nuclear fusion[1] and aviation. The highest dissipation-less current is named the critical current, and this is one of the prime practical properties of superconductors (together with the critical current density, $J_c$). Recently, Goyal *et al*[2] reported a record high $J_c$~190 MA/cm$^2$ at 4.2 K in (RE)BCO films, exceeding the highest $J_c$ in the best commercial (RE)BCO wires by a factor of five. Based on the huge potential practical impact of this high $J_c$, we examined the raw experimental data[2] and found that this high value originates from an error in the conversion of units. The real $J_c$ is 10 times smaller than the reported $J_c$, consistent with values currently achieved by many manufacturers.**

In Ref.[3] we established the fundamental nature of the self-field critical current density, $J_c(\text{sf},T)$, in superconductors as a fundamental materials parameter which for thin-film superconductors is defined by the equation:

$$J_c(\text{sf},T) = B_{c1}(T)/\mu_0\lambda(T) = \frac{\phi_0}{4\pi\mu_0} \times \frac{\ln(\lambda/\xi)+0.5}{\lambda^3(T)} \quad (1)$$

where, $B_{c1}(T)$ is the lower critical field, $\phi_0$ is the magnetic flux quantum, $\mu_0$ is the permeability of free space, $\lambda(T)$ is the London penetration depth, and $\xi(T)$ is the coherence length. This relation was demonstrated eventually for over 100 superconductors in the thin limit.

The notable feature of Eq. 1 is that it established an upper limit for the self-field transport critical current in thin-film superconductors based on fundamental material parameters, where the dominant role is played by the London penetration depth, or alternatively, by the density of Cooper pairs in the condensate. Here, the influence of another fundamental length of superconductors, the coherence length, is very weak.

Importantly, from a practical standpoint, this fundamental limit should apply to the so-called second-generation high-temperature superconducting wires (HTS 2G-wires or $REBa_2Cu_3O_{7-\delta}$ tapes), which are the preferred option for high-field applications such as magnets for fusion projects[1].

Accordingly, the highest $J_c$(sf,$T$) would be achieved with the lowest possible value of $\lambda$, and in $REBa_2Cu_3O_{7-\delta}$ this can only be achieved by full oxygenation ($\delta \rightarrow 0$) and removing all impurity substitution. Based on the lowest reported values[4] for $REBa_2Cu_3O_7$ of $\lambda_a$ = 103 ± 8 nm and $\lambda_b$ = 80 ± 5 nm, giving an effective in-plane penetration depth of $\lambda_{eff} = \sqrt{\lambda_a \lambda_b}$ = 91 nm, combined with a value $\xi_0$ = 1.5 nm, this limit should be $J_c(\mathrm{sf}, T \rightarrow 0\ \mathrm{K}) \leq 78$ MA/cm$^2$. Such a figure has been reported by Stangl et al.[5], for overdoped pulsed-laser-deposited films, and most major 2G-wire manufacturers around the world are now within a factor of two or three of this limit[6]. And, despite the fact that nanoengineering can be used to enhance the *in-field* $J_c$ in $REBa_2Cu_3O_{7-\delta}$, this fundamental *self-field* $J_c$ is unlikely to be exceeded.

Hence our surprise to note the recent claim by Goyal *et al.*[2] of $J_c(B=0, 4.2\ \mathrm{K})$ = 190 MA/cm$^2$ reported for a $Y_{0.5}Gd_{0.5}Ba_2Cu_3O_x$+2%$BaZrO_3$ film, exceeding our proposed upper limit by a factor of nearly three.

It should be noted that the only measurement relevant to industry is the *transport* $J_c$, arising from passing a current through the conductor[7]. However, Goyal *et al.*[2] measured only the *magnetization* $J_c$ (designated below as $J_{c,mag}(B,T)$) which is calculated from the measured magnetic moment, $m(B,T)$, of the sample on completing a full cycle of the magnetic hysteresis loop, $m$ versus applied field, $B$.

The authors, Goyal *et al.*[2], kindly provided us with raw measured $m(B,T)$ datasets for three samples shown in their Figs. 1, 2 and S6 measured at $T$ = 10, 30, 65, 77 K.

We calculated $J_{c,mag}(B,T)$ from $m(B,T)$ using the standard textbook procedure (see, for instance, Eq. 13.30 in Poole *et al.* (2007)[8]) and research papers[9–13] for a superconductor having the form of a rectangular prism:

$$J_{c,mag}(B,T) = 4 \times \frac{m(B,T)}{w \times l \times t} \times \frac{1}{w \times (1 - w/3l)} \qquad (2)$$

where $m(B,T)$ [Am$^2$] is the measured irreversible magnetic moment of the sample in its saturated state at applied field $B$, $w$ is sample width, $l$ is sample length, and $t$ is sample thickness, where $t \leq w \leq l$, and the dimensions are in [m]. Because all values in Eq. 2 are in SI units, we refer to Eq. 2 as the "*standard SI equation*". Because magnetometers usually generate magnetic moment $m(B,T)$ in units of [emu], it is useful to keep in mind the conversion 1 [emu] $\rightarrow$ 0.001 [Am$^2$].

Our central finding is that our calculations result in $J_{c,mag}$ values 1/10$^{th}$ those reported by the authors[2]. For instance, in sample $Y_{0.5}Gd_{0.5}Ba_2Cu_3O_x$+2%$BaZrO_3$ (shown in Figs. 2 and 9 of ref.[2]) the authors reported $J_{c,mag}(B=0, 10\ \mathrm{K})$ = 185 MA/cm$^2$, while our calculations showed $J_{c,mag}(B=0, 10\ \mathrm{K})$ = 18.5 MA/cm$^2$.

In contrast to the standard Eq. 2, the authors[2] used the equation proposed by Gyorgy *et al.*[14]:

$$J_{c,mag}(B,T) = 20 \times \Delta M \times \frac{1}{w \times (1 - w/3l)} \qquad (3)$$

where the current density $J_{c,mag}$ is in [A cm$^{-2}$], $\Delta M = \frac{m(B_+,T) - m(B_-,T)}{w \times l \times t}$ is in [emu cm$^{-3}$] (i.e. in gaussian units) and this is the unit-volume full-width of the hysteresis loop for increasing and decreasing applied magnetic field. The dimensions are in [cm].

Crucially, the multiplicative term, 20, is not dimensionless. Examination of the units of the left- and right-hand sides of Eq. 3 reveals that its units are [A cm$^2$ emu$^{-1}$]. However, Goyal *et al.*[2] converted the measured magnetic moment $m(B,T)$ from units of [emu] into units of [A cm$^{-1}$] by the following protocol:

$$m(B_+,T) - m(B_-,T)\ [\text{emu}] \rightarrow [m(B_+,T) - m(B_-,T)]/(w\times l\times t)\ [\text{emu cm}^{-3}]$$

$$\rightarrow 10 \times [m(B_+,T) - m(B_-,T)]/(w\times l\times t)\ \ [\text{A cm}^{-1}]$$

$$= \Delta M\ \ [\text{A cm}^{-1}] \qquad (4)$$

The converted $\Delta M$ [A cm$^{-1}$] values (using Eq. 4) are substituted in the right-hand side of Eq. 3, from which $J_c$ is calculated. This procedure (Eqs. 3,4) always results in a 10-fold error in comparison with the real $J_c$. The reason for this 10-fold error is that Eq. 3[14] had already been simplified to eliminate all units conversions. Thus, despite the fact that the conversion of measured $m(B,T)$ [emu] data into $\Delta M$ [A cm$^{-1}$] using Eq. 4 is formally correct, Eq. 3 does not intend that this conversion should be implemented, because Eq. 3 already contains the conversion.

To demonstrate this our primary message, in Figure 1 we analyzed $m(B,T$=18 K) data reported by Naqib and Islam[13] for a $Y_{0.90}Ca_{0.10}Ba_2Cu_3O_{7-\delta}$ film with $w$ = 5 mm, $l$ = 10 mm, $t$ = 280 nm. The conversion of the measured $m(B,T$=18 K) values into $J_{c,mag}(B,T$=18 K) by the standard procedure (Eq. 2) is shown in Fig. 1,b while in Fig. 1,c we show $J_{c,mag}(B,T$=18 K) calculated by the approach implemented by Goyal *et al.*[2] (i.e. by Eqs. 3,4). The latter approach gives $J_{c,mag}$(0, 18 K) = 184 MA/cm$^2$, which betters the value $J_{c,mag}$(0, 10 K) = 180 MA/cm$^2$ reported by Goyal *et al*[2] as "the highest values of $J_c$ obtained to date".

Similarly, the correct value for the $(Y_{0.5}Gd_{0.5})Ba_2Cu_3O_x$ film (shown in the authors' Fig. 1) is $J_{c,mag}$(0, 10 K) = 4.4 MA/cm$^2$ vs the reported value[2] of $J_{c,mag}$(0, 10 K) = 44 MA/cm$^2$. Moreover, for the $(Y_{0.5}Gd_{0.5})Ba_2Cu_3O_x$+2at.%BZO film (shown in Figure S6[2]) the authors again reported[2] a value of $J_c$ 10 times higher, $J_{c,mag}$(0, 10 K) = 175 MA/cm$^2$, than the true value of $J_{c,mag}$(0, 10 K) = 17.5 MA/cm$^2$.

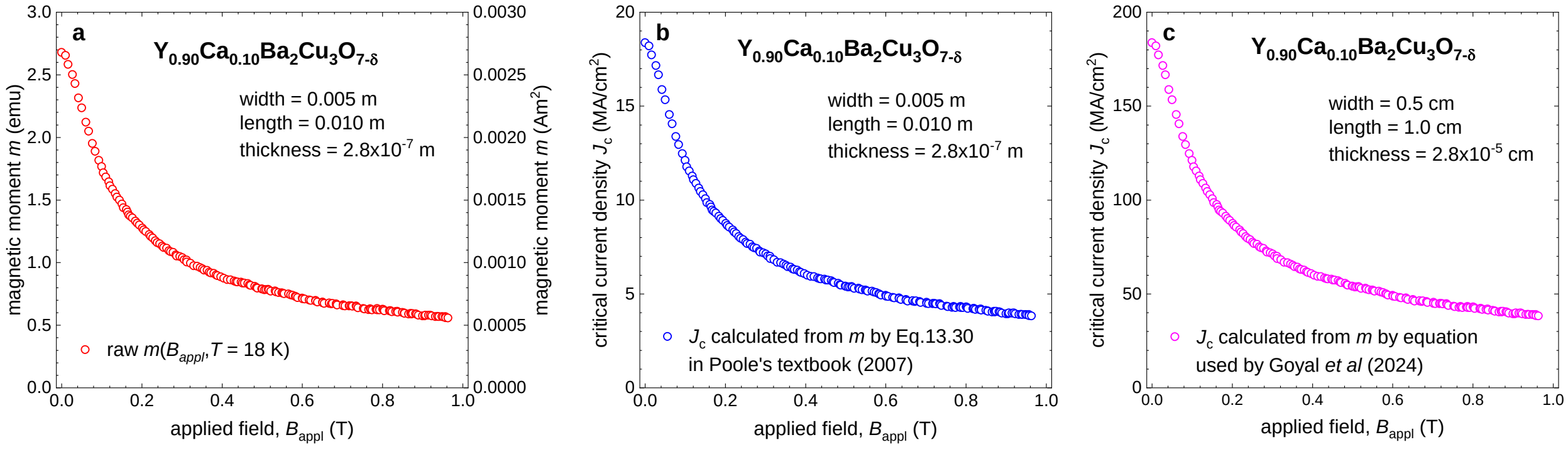

**Figure 1.** (a) Measured magnetic moment $m(B,T$=18 K) for a $Y_{0.90}Ca_{0.10}Ba_2Cu_3O_{7-\delta}$ film with $w$ = 5 mm, $l$ = 10 mm, $t$ = 280 nm, and the conversion of the moment into $J_{c,mag}(B$, 18 K) by (b) standard textbook SI routine (Eq. 2) and (c) by the approach implemented by Goyal *et al.*[2] and described by Eqs. 3,4. The raw $m(B$,18 K) dataset was reported by Naqib and Islam[13].

For final confirmation of our primary message, we performed measurements of the magnetic moment $m(0,T)$ for an S-Innovations (formerly SuperOx) $YBa_2Cu_3O_{7-\delta}$ 2G-wire with a superconducting layer of thickness $t = 2.0$ μm. We compared the calculated $J_{c,mag}(0, T)$ values with the transport critical current density, $J_{c,tr}(sf, T)$, measured for a similar $YBa_2Cu_3O_{7-\delta}$ tape fabricated by the same manufacturer which has a thickness of the superconducting layer of $t$ = 2.82 μm. These transport measurements are available online[6].

Results are shown in Fig. 2, where one can see that $J_{c,tr}(B,T) \approx J_{c,mag}(B,T)$ when the latter is calculated by Eq. 2 (Fig. 2,b). However, if $J_{c,mag}(B,T)$ is calculated by the approach used by Goyal *et al.*[2], then it is about 10 times larger than $J_{c,tr}(B,T)$ (Fig. 2,c).

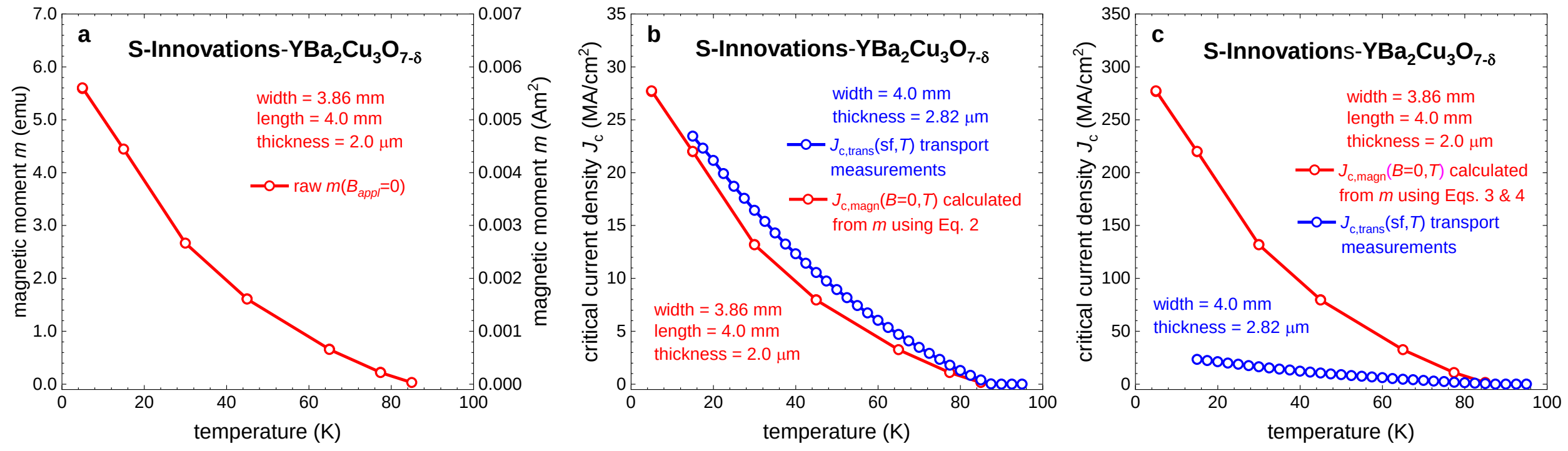

**Figure 2.** (a) Measured magnetic moment $m(0,T)$ for the S-Innovations (formerly SuperOx) $YBa_2Cu_3O_{7-\delta}$ tape with $w$ = 3.86 mm, $l$ = 4 mm, $t$ = 2.0 μm. (b) $J_{c,mag}(0,T)$ (calculated by Eq. 2 from $m(0,T)$ in Fig. 2,a) with transport $J_{c,tr}(sf,T)$ measured for another S-Innovations $YBa_2Cu_3O_{7-\delta}$ 2G-wire with dimensions shown in the panel[6]. (c) $J_{c,mag}(0,T)$ (calculated from $m(0,T)$ in Fig. 2,a by the approach used by Goyal *et al.*[2] (Eqs. 3 and 4) and compared with $J_{c,tr}(sf,T)$.

We have identified other '10-fold mistakes' in the paper[2]. For instance, in the Williamson-Hall (WH) analysis[15] shown in Fig. 3,c[2], one can see that the slope of the WH line is 0.00201, however the authors wrote[2]: "*For the* $(Y_{0.5}Gd_{0.5})Ba_2Cu_3O_X$ *film, residual microstrain was estimated to be 2.01%.*" In Fig. 4,c[2] the slope of the WH line is 0.00266, however the authors wrote[2]: "*The residual microstrain was estimated to be 2.66%.*"

In summary, we have shown that the record high $J_c$ reported by Goyal *et al.*[2] originated from an error in the conversion of units. A correct analysis shows 10 times smaller $J_c$ values, consistent both with the all-important transport $J_c$ values and well below values currently achieved by many manufacturers. In the process, we have also confirmed the validity of our primary equation (Eq. 1) for the self-field critical current density in superconductors[3].

## Declaration of competing interest

The authors declare that they have no known competing financial interests or personal relationships that could have appeared to influence the work reported in this article.

## Data availability

Data that supports this study is available from authors upon request.

## Acknowledgements

EFT thanks financial support provided by the Ministry of Science and Higher Education of Russia (theme 'Pressure' No. 122021000032-5).

## Contributions

Both authors contributed equally.

## Corresponding authors

Correspondence to E.F. Talantsev (evgeny.f.talantsev@gmail.com) and J.L. Tallon (Jeff.Tallon@vuw.ac.nz)